\documentclass[prl,twocolumn,letterpaper,amsfonts]{revtex4}
\usepackage{graphicx}

\usepackage{amsthm}
\usepackage{amsmath}
\usepackage{amsfonts}

\usepackage{amssymb}

\usepackage{bm}

\begin{document}

\bibliographystyle{apsrev}

\newcommand{\bra}[1]{\ensuremath{\left \langle #1 \right |}}
\newcommand{\ket}[1]{\ensuremath{\left | #1 \right \rangle}}
\newcommand{\braket}[2]{\ensuremath{\left \langle #1 \right | \left. #2 \right \rangle}}
\newcommand{\tr}{\ensuremath{\mbox{Tr}}}
\newcommand{\R}{{\bf r}H}
\newcommand{\V}{\widehat{{\bf r}H}}
\newcommand{\U}{{\bf t}H}
\renewcommand{\H}{\ensuremath{\mathcal{H}}}
\newcommand{\var}{\ensuremath{ \mbox{var}}}
\newcommand{\hvar}[1]{\ensuremath{ \hat{\mbox{var}} \left ( #1 \right )}}
\newcommand\Prefix[3]{\vphantom{#3}#1#2#3}

\title{Using the Memories of Multiscale Machines to Characterize Complex
Systems}
\author{Nick S. Jones}

\affiliation{Oxford Centre for Integrative Systems Biology, Departments of Physics and Biochemistry, Oxford University, 
Oxford, UK}


\begin{abstract}

A scheme is presented to extract detailed dynamical signatures from successive measurements of complex systems. Relative entropy based time series tools are used to quantify the gain in predictive power of increasing past knowledge. By lossy compression, data is represented by increasingly coarsened symbolic strings. Each compression resolution is modeled by a machine: a finite memory transition matrix. Applying the relative entropy tools to each machine's memory exposes correlations within many timescales. Examples are given for cardiac arrhythmias and different heart conditions are distinguished.
%
%
%

\end{abstract}

\maketitle

Before understanding a complex system, one often needs to
interpret the complex signals it generates.
%
%
While it is easy to find correlations between arbitrarily separated pairs of points in a signal time series \cite{Voss}, highly correlated signals can lack such pairwise correlations (Section I).
%
Instead, at a higher resolution, one can estimate the joint probabilities of sequences of events and, eg., find the order of corresponding Markov chains \cite{Billingsley}. Unfortunately the set of possible event sequences will generically increase exponentially with sequence length while becoming proportionately harder to estimate; i.e. what these methods gain in resolution over
pairwise statistics, they lose in range. We must, however, expect Nature to show correlations which are both long-ranged and more than pairwise.

 This Letter suggests a tool, akin to
the autocorrelation, which is intuitive, sensitive to more than
pairwise correlations and yet is long-ranged enough to capture the
longtime correlations shown by some complex systems. It combines
two core ideas: 1) a natural measure of the predictive power one
gains as one has an increasingly long symbolic string (Sec. II-V);
2) use of  lossy compression to express the dynamics of a complex
system as a set of Markov sources (transition matrices) with each
one representing the dynamics on a different timescale (VI-VIII). %
The following considers a system which, at any time $t$, can be in
state $x_t$ chosen from an alphabet (finite set)
$\mathcal{A}$. The system passes through states $x_1,x_2...x_T$ at
fixed intervals and the data is ergodic and stationary. From now on
$x_j,...x_k$ will be represented by $x^k_j$. Having addressed
pairwise correlation measures in Section I, Sections II-VI develop a
new means of mapping correlations in strings and VII-IX consider
physiological examples and continuous time series.
%
%
%

\emph{I Pairwise Statistics Can Fail to Capture Structure.} Any
approach which investigates the time structure of a data string must
be compared with conventional methods like the
 autocorrelation function. 
 Pairwise measures, which compare a symbol at one point in a
 string with a (possibly different) symbol at another point, fail to
 capture conditional behavior on other intervening symbols. The following one-parameter, order-two transition matrix 
 with 
 alphabet $\mathcal{A}_4=\{A,B,C,D\}$ should remind the reader of this
 phenomenon; it can create strings without pairwise correlations. Using the notation that $X$ is a random variable and
$x$ is a particular instantiation of that variable, it is of the
form
 $p(X_3|X_1X_2)$ with ($x_i\in\mathcal{A}_4)$:

{\small

\begin{equation}\label{8.5}
\begin{tabular}{|l|l|l|l|l|l|} \hline
    $\mbox{\tiny{\emph{x}}}_{\mbox{\tiny{\emph{1}}}}{\mbox{\tiny{\emph{x}}}_{\mbox{\tiny{\emph{2}}}}\backslash^{{x}_{\it 3}}}$&&$A$ &$B$ &$C$ & $D$ \\\hline\hline
$AA$&&$\frac{1}{4}\;\;$ &$\frac{1}{4}\;\;$ &$\Lambda^+$ &
$\Lambda^-$
\\\hline
$AB$&&$\frac{1}{4}\;\;$ &$\frac{1}{4}\;\;$ &$\Lambda^-$ &
$\Lambda^+$
\\\hline
$BA$&&$\frac{1}{4}\;\;$ &$\frac{1}{4}\;\;$ &$\Lambda^-$ &
$\Lambda^+$
\\\hline
$BB$&&$\frac{1}{4}\;\;$ &$\frac{1}{4}\;\;$ &$\Lambda^+$ &
$\Lambda^-$
\\\hline
$pq$&&$\frac{1}{4}\;\;$ &$\frac{1}{4}\;\;$ &$\frac{1}{4}\;\;$ &
$\frac{1}{4}\;\;$
\\\hline
\end{tabular}
\end{equation} }

where $\Lambda^{\pm}=\frac{1}{4}\pm d$, the free parameter $d$ is
$0\leq d\leq \frac{1}{4}$ and where $pq$ is any ordered pair, $p$
preceding $q$, other than $AA,BB,BA,AB$. One can prove that, for
strings generated by this matrix, the probability of
 obtaining the symbol $u\in\mathcal{A}_4$ an interval $l\neq 0$ after
 $v\in\mathcal{A}_4$ is independent of both $l$ and $v$. 
 The
 auto/cross correlations of such a series are thus
 indistinguishable from white noise (for explanations of symbolic autocorrelations see Voss \cite{Voss}). However,  the string is highly
 structured; measures like $\R_{n||m}$ below can  expose
 this. Simple measures which reveal
 correlations between points, without having to create data objects
 which scale exponentially with order, can be very useful. 
 The approach in Section VI yields data objects that both increase slowly with order/time and  illuminate more than pairwise correlations.

 \emph{II The Transition Entropy.} If a string is sufficiently long,
one can estimate transition probabilities $p(X_{t}|X_{t-m}^{t-1})$.
It should be stressed that this paper is not, in the first instance,
about the estimation of these probabilities; we will assume that
they are given to us exactly \cite{Herzel94}. The following
entropies investigate the structure of this order $m$ transition
matrix.

%
%
Call $\small {H(X_t)=\sum_{x_t\in \mathcal{A}}
-p(x_{t})\log_{\tiny{2}}p(x_{t})}$ the Shannon entropy of $X_t$. The
transition, or conditional, entropy $\U_m$ is defined as follows:
{\small
\begin{eqnarray}
\U_m&=&\sum_{x^{t-1}_{t-m}\in \mathcal{A}^m}p(x^{t-1}_{t-m})H(X_t|x^{t-1}_{t-m}),\label{avHT}
\end{eqnarray}\small} 
 This transition entropy, $\U_m$, measures the entropy of
predictions one makes when equipped with a length $m$ string, when
one does not know what the string is. 
If each length $m$ string that occurs exactly predicts the next
state, then $\U_m=0$ (for the series $...ABABA...$ each length $1$
string uniquely determines its ensuing state: $\U_1=0$). If no
 string imparts a
predictive advantage then $\U_m=H(X_t)$ $\forall m$. 
Convexity arguments \cite{Berger} show that
$\U_m\geq \U_n$ ($n>m$).
%

 \emph{III The Relative Transition Entropy.}
The \emph{relative transition entropy}, $\R_{n||m}$,  defined below
is a measure of gain in predictive power as one moves from knowledge
of a length $m$ string to a length $n$ string ($n>m$). The relative
entropy or Kullback-Leibler divergence \cite{Berger} between the
distribution $Q(X)$ and $P(X)$, where $X$ can take $|\mathcal{A}|$
different values, is: {\small{$D(Q||P)=\sum_{x\in\mathcal{A}}Q(x)\log\frac{Q(x)}{P(x)}$}}. 
It is often described as the average disbelief in a
model's predicted distribution $P$, when observing random outcomes $X$ from real data $Q$. 
%
By contrast, we will use it to capture the degree that predictions
made when equipped with more knowledge of the past, represented by
$Q$, are inconsistent with those made with reduced knowledge, $P$.
%

One can compare predictions about a symbol at time $t$ given
knowledge of a particular set of preceding $n$ symbols
($x^{t-1}_{t-n}$) with predictions made when given only the
preceding $m$ ($x^{t-1}_{t-m}$ with $n>m$). The divergence,
$D(p(X_{t}|x^{t-1}_{t-n})||p(X_{t}|x^{t-1}_{t-m}))$,
 measures the
information lost if one loses the knowledge that the sequence
$x^{t-1}_{t-m}$ was preceded by $x^{t-m-1}_{t-n}$. Averaging over
all strings $x^{t-1}_{t-n}$ yields the
 relative transition entropy $\R_{n||m}$: {\small
\begin{equation}
\R_{n||m}=\sum_{x^{t-1}_{t-n}}p(x^{t-1}_{t-n})D(p(X_{t}|x^{t-1}_{t-n})||p(X_{t}|x^{t-1}_{t-m})).
\label{avHTR}
\end{equation}}
This quantifies the predictive power lost when one moves from having
a length $n$ string of prior information to the shorter length $m$,
for a randomly selected string $x^{t-1}_{t-n}$ \cite{Schreiber}. Let us now establish
a few properties of $\R_{n||m}$. Using (\ref{avHT}-\ref{avHTR}) one
can readily prove that $\R_{n||m}=\U_{m}-\U_n$,  $n>m$.
 If $\U_{m}=\U_n$ then 
$\R_{n||m}=0$. Since $D(Q||P)\geq0$, with equality only when $Q=P$,
we further know that if $\R_{n||m}=0$ then
$p(X_{t}|x^{t-1}_{t-n})=p(X_{t}|x^{t-1}_{t-m})$ $\forall X_{t},\;
x^{t-1}_{t-n}$. The length $n$ and $m$ predictions are
\emph{exactly} the same. In general $D(Q||P)$ can be unbounded
\cite{Berger}, but here, some thought shows that $0\leq
\R_{n||m}\leq H(X_t)$. We can now formulate a hierarchy of
differential quantities. Defining the Shannon entropy for strings of
length $n$ as $H_n=H(X^t_{t-n+1})$, one readily finds that
$\U_n=H_{n+1}-H_{n}$ and $\R_{n||m}=\U_{m}-\U_n$. Authors have noted
that the way that $H_n$ and $\U_n$ decrease with $n$, reveals
structure in the string \cite{grassbergertending,Bandt}: in Section
VI we will use $\R_{n||m}$ to map these correlations
\cite{Schreiber}.

\emph{IV Example: $\R_{n||m}$ for the distribution in Eq.
\ref{8.5}}.  $p(X_3|X_1X_2)$ yields the stationary state,
$p(X_1X_2)=\frac{1}{16}$ so $p(X_2|X_1)=\frac{1}{4}$. By Eq.
\ref{avHTR} one finds $\R_{1||0}=0$ (comparing $p(X_2|X_1)$ and
$p(X_2)$). Comparing $p(X_3|X_1X_2)$ and $p(X_3|X_2)$ shows that
$\R_{2||1}=\frac{1}{4}\log_2[2(\Lambda^+)^{\Lambda^+}(\Lambda^-)^{\Lambda^-}]$.
I.e. knowing the current state is of no help in predicting the next
state ($\R_{1||0}=0$) but knowing the current and preceding state
does help ($\R_{2||1}>0$ and so $\R_{2||0}>0$). Since, for
$n>2,m>1$, $\R_{n||m}=0$
 one concludes that knowing more than the preceding and current state gives
  no further predictive advantage.

\emph{V Introducing a measure to detect concealed structure.} We
noted that if the predictions of length $m$ and $m+1$ strings differ
then $\R_{m+1||m}>0$; however, since Eq. \ref{avHTR} is an average,
small changes in this quantity can hide dramatic changes between the structure of order $m+1$ and order
$m$ transition matrices. One can readily construct examples where
there exists a string $x^{t-1}_{t-m-1}$ such that
$D(p(X_{t}|x^{t-1}_{t-m-1}||p(X_{t}|x^{t-1}_{t-m}))\gg
\R_{m+1||m}$.  It is thus useful to introduce the
quantity: 
$\small{\V_{m+1||m}=\max_{x^{t-1}_{t-m}}D(p(X_{t}|x^{t-1}_{t-m-1}||p(X_{t}|x^{t-1}_{t-m})))}$
the \emph{maximum relative transition entropy} over all strings
$x^{t-1}_{t-m}$ in $A^m$.
This measures when knowledge of a \emph{particular} extra symbol
imparts a large predictive advantage.

\emph{VI Introducing Multiscale Markov Sources.} This section
introduces a method for describing data from complex systems by
fitting finite state machines with memories to each of their
different time-scales. Consider coarsening time series to lower and
lower time resolutions. For each resolution one might estimate a
small, order $m$, transition matrix (Markov source is another name
for transition matrix \cite{Berger}). Let us call these matrices,
one for each resolution, a set of multiscale Markov sources. Suppose
the real data was generated by a high order Markov source of order
$l\gg m$. An order $l$ source (alphabet $\mathcal{A}$) has
$|\mathcal{A}|^{l}(|\mathcal{A}|-1)$ parameters. By contrast, we
will see that a corresponding set of multiscale Markov sources
requires only $\sim\log l$ parameters. The sources thus form a
compact multiscale representation of the data.

Let us now examine more details of the coarsening. We first break the symbolic series, $y^T_1$, $y_i\in\mathcal{A}$,  into consecutive non-overlapping blocks, each $c^r$ symbols long. We fix $c$, the basic block size, and let $r$ vary to give different block sizes $c^r$ (increasing $r$ increases the block size and we will see that this lowers the resolution).  Then we
coarsen by mapping each possible block (of which there are
$|\mathcal{A}|^{c^r}$) onto a single symbol from a smaller set
$\mathcal{C}$ ($|\mathcal{C}|<|\mathcal{A}|^{c^r}$). The manner of
this map will be discussed below. The new coarsened string at
resolution $r$ has ${T/c^r}$ elements $^rx^{T/c^r}_1$ with $^rx_i\in
\mathcal{C}$. From this string one can estimate an order $m$ Markov
source, $^r_mM$. Supposing the raw data was generated by a source of
order $l$, one might fix $|\mathcal{C}|$, $c$ and $m$ and vary $r$
to give a set of Markov sources $^r_mM$
 with $r\in\{1,2,...\lceil \log_c \frac{l}{m}\rceil\}$. By choosing
 this range of $r$ values
 the set of sources has a similar memory to
the order $l$ transition matrix. While the order $l$ source needs
$|\mathcal{A}|^{l}(|\mathcal{A}|-1)$ parameters, the total number of
parameters in the multiscale sources is
$|\mathcal{C}|^{m}(|\mathcal{C}|-1) \lceil log_c \frac{l}{m}\rceil$.
The set of Markov sources $^r_mM\;\forall r\leq \lceil \log_c
\frac{l}{m}\rceil$ thus gives a compact multiscale dynamic model for
the correlations at each timescale \cite{Dugatkin} (see top diagram
in Fig. \ref{threehearts}).

Such lossy 
compression lies  broadly within
rate-distortion theory  \cite{Berger}. 
Distortion measures capture the lossiness of maps from blocks to single
symbols. 
The following motivational example uses the
crude Hamming distortion. Blocks of $c^r$ symbols, each symbol
in $\mathcal{A}$, can be viewed as coming from an alphabet
$\mathcal{B}_{c,r}$ of size $|\mathcal{A}|^{c^r}$. Call a
compression a map $f: \mathcal{B}_{c,r}\rightarrow \mathcal{C}$.
A map is optimal
if a version reconstructed from the compressed string (using an
inverse map $g: \mathcal{C} \rightarrow
 \mathcal{B}_{c,r}$) and the original string are as
close as possible with respect to a given measure. The Hamming
distortion is $d(g(f(X_t)),X_t)=\delta_{g(f(X_t)),X_t}$ for
$X_t\in\mathcal{B}_{c,r}$. Given $p(X_t)$,  the optimal map
 minimizes the expected symbol-by-symbol distortion between
 the reconstructed and original letters $<d(g(f(X_t)),X_t)>$. 
 Here, some thought shows that the optimal
 $f$: (1) takes each of the $|\mathcal{C}|$ most probable symbols in $\mathcal{B}_{c,r}$ to a distinct
 symbol in $\mathcal{C}$  and (2) takes all other symbols to an arbitrary symbol in $\mathcal{C}$. 
%

\begin{figure}[h!]
\includegraphics[angle = 0, width = 8.3cm,
keepaspectratio=true]{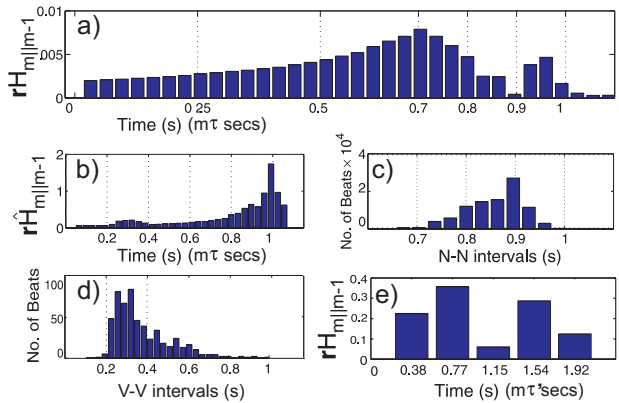}
\caption{\label{scaling3547} A patient with cardiac arrhythmia: a) 
The  $m^{\rm{th}}$ bar gives the predictive advantage of knowing
$m\tau$ seconds of past activity over knowing $m\tau-\tau$ seconds of activity 
($\R_{m||m-1}$).
$\tau=32$ms and $t\leq35\tau$ 
(more than $99\%$ of all length $35\tau$ strings, picked uniformly
at random from the data, occurred more than $300$ times). b)
$\V_{m||m-1}$ for the original data. c,d) The
intervals between successive normal and ventricular beats
respectively. e) $\R_{m||m-1}$ for a
coarsened string $r=1$, $c=12$, $|\mathcal{C}|=10$ each timestep is
thus $\tau'=0.384$s. For length 4 and 5 strings in the coarsened
data, $97\%$ and $90\%$ respectively of all such strings picked
uniformly at random occurred more than $300$ times.}
\end{figure}

\emph{VII Example: Sudden Cardiac Death.} This section applies the
above tools to heart arrhythmia. A simplified view of the heart is
that, in any short interval, it can have a normal ($N$) or
ventricular beat ($V$) or no beat at all ($\O$). The raw data is a
list of times of beats labeled as $N$ or $V$ \cite{physionet}. By
discretizing time into blocks of $\tau=32$ms this list was converted
into a symbolic string of the form `... $\O N\O V\O\O N$...'. The
following uses 24 hours of heart data for a patient with many $V$
beats. Given this three-letter alphabet one can attempt to estimate
a transition matrix of order $m$: $p(X_{t}|x^{t-1}_{t-m})\; \forall
X_{t}, x^{t-1}_{t-m}$ \cite{Courtemanche}. Since $\tau\ll$ interbeat
interval ($\sim0.85$s) and the system is very structured, the size
of the transition matrix grows slowly with $m$. Accommodation of
finite size effects in the estimation of such transition matrices,
is delicate  \cite{Herzel94} and a  crude approach was used here
(partly justified by the wealth of data). The transition matrices
can only give information about the data as a whole (rather than the
behavior of the heart at any one time) as, alongside the presence of
multiscale nonstationarities \cite{multiscale}, this patient was
particularly unhealthy. Fig. \ref{scaling3547}a) shows $\R_{m||m-1}$
for the patient.  As $m\tau$ (the duration of string one is given)
increases towards the $N-N$ beat interval (see Fig.
\ref{scaling3547}c)), one's ability to make good predictions
increases markedly. But, when $m\tau$ nears the heart beat interval,
further knowledge gives less predictive advantage (because one is
already equipped with knowledge of a characteristic time period of
the process). As a result $\R_{m||m-1}$ begins to fall around
$0.7$s, mirroring the distribution in Fig. \ref{scaling3547}c).
Beats with intervals $>1s$ are rare  so $\R_{m||m-1}$ for $m\tau>1$s
is small.  Fig
\ref{scaling3547}b) plots $\V_{m||m-1}$ (with the strong promise
that all strings considered occurred more that $300$ times in the
data). It reveals hidden structure between $0.2$ and $0.4$s. This
peak is the compound effect of short $V-V$ events and misannotations
in the
uncorrected  record (see Fig. \ref{scaling3547}d)). The coarsened data, Fig. \ref{scaling3547}e), reveals
structure on another timescale: one sees that a large part of the
predictive knowledge is contained in the first second of activity
but another characteristic timescale, open to physiological interpretation, appears in the range $1.19-1.54$s. 
Plots like Fig. \ref{scaling3547} might distinguish between
 heart conditions, since 
these can depend on dynamics of a few seconds
\cite{Schulte}.

\begin{figure}[h!]
\includegraphics[angle = 0, width = 6.5cm, 
keepaspectratio=true
]{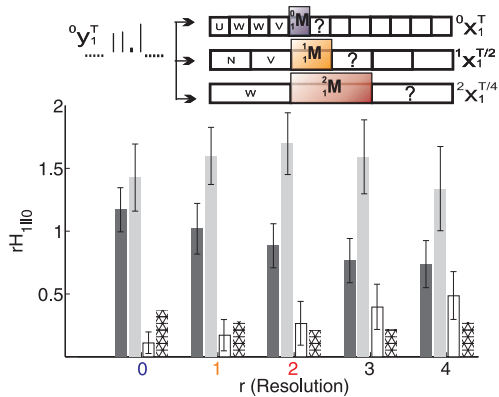}
\caption{\label{threehearts}
  \emph{Above}: schematic of vector quantization (${}^0y^T_1\rightarrow {}^rx^{{T}/{c^r}}_1$) and multiscale Markov sources ${}^r_1M$.
  \emph{Main}: The black, grey and white bars are the means, $\pm$ standard errors, of groups of patients who
  were healthy, experiencing congestive heart failure
and atrial fibrillation respectively. Unphysiological and ectopic
beat intervals were filtered  (as in \cite{Costa}) and hourly trends removed. The fourth,
hatched, columns are for random phased $1/f$ noise. The $x$ axis
considers the beat interval data at five different resolutions
$r=0...4$ $c=2$. At each resolution $rH_{1||0}$ is found; this
measures the extra predictive power from knowing one symbol over
knowing none. Gaussian white noise has $rH_{1||0}\rightarrow 0\;
\forall r$ and is not plotted (Brownian noise has
$rH_{2||1}\rightarrow 0\; \forall r$) .
}
\end{figure}

\emph{VIII Continuous time series.} Multiscale Markov sources can also be found for continuous time series (eg. wind speed) as well as symbolic strings. The
data, sampled at $T$ times, is again broken up into blocks of $c^r$
consecutive points and coarsened. The alphabet we compress to is now
a set of $|\mathcal{C}|$ letters with
each one representing a different motif of $c^r$ consecutive reals. 
We allocate each block of raw data to its closest motif using a
mean-square distance.
%
%
For example, suppose we want to compress blocks of two data points
$(c=2,r=1)$ to one of three symbols and we are given a three letter
code book with three motifs: $\mathcal{C}=\{N=(1,1),\, V=(0.1,2),\,
U=(10,10)\}$. Using the mean-square distance, the sequence of
continuous data $^{0}y^T_1=...|1\; 1.1|0.2\; 1.5|1.1\; 1.01|...$ is
optimally represented by
$^{1}x^{{T}/{2}}_1=...|N|V|N|...$ See Fig \ref{threehearts}. 
The optimal set of motifs, for a fixed $|\mathcal{C}|$, allow the
compressed sequence to reconstruct the original with minimum total
mean-square error \cite{costacomm}. 
A selection of algorithms exists for finding optimal motifs (vector
quantizers \cite{Berger}). In our case, such algorithms have as
input the set of $T/c^r$ blocks of length $c^r$ and the value of
$|\mathcal{C}|$. They output the set,  $\mathcal{C}$, of motifs of
length $c^r$ which minimizes the total mean-square error for this
block size. Given these motifs one can then convert the continuous
time series into its closest symbolic equivalent (see the above
example: $^{0}y^T_1$ $\rightarrow {}^{1}x^{{T/2}}_1$ see Fig.
\ref{threehearts}). Given this string of symbols generated from
blocks of $c^r$ real valued data points, one again determines
$^r_mM\;\forall r$.

\emph{IX Example.} Fig. \ref{threehearts}
applies this idea to three different groups of cardiac patients
\cite{physionet}, using the Generalized Lloyd algorithm to find the
appropriate set of motifs, $\mathcal{C}$, for each resolution, $r$
\cite{Berger}. The raw cardiac data was an `interval series': each
data point being the time interval between successive heart beats.
The matrix $^r_1M$ was found for $r=0...4$, $c=2$,
$|\mathcal{C}|=10$ and the extra predictive power from
knowing one symbol was estimated: $rH_{1||0}$. 
The three different heart conditions can be distinguished (for
comparable results see \cite{Costa}); healthy hearts show a slow
loss in predictability, the disordered beats that occur in atrial
fibrillation yield a low degree of predictability whereas congestive
heart failure shows an increase in predictability at some scales.

\emph{X Conclusion.} 
%
%
%
This paper presented a means of producing a one parameter map of
predictive knowledge acquired as one is equipped with increasingly
long substrings of a symbolic data set. It suggests that lossy
compression allows this short range mapping technique to be
compactly extended to the study of longer ranged correlations.
Examples are given where different heart conditions are
distinguished and characterized using these methods. Underlying this
work is the view that dynamical signatures of some systems can be
found by treating them as sets of Markov sources with each source
characterizing dynamics on a different timescale.

Thanks to M Costa, A Goldberger, C-F Lee and C-K Peng

%


\begin{thebibliography}{30} \expandafter\ifx\csname
natexlab\endcsname\relax\def\natexlab#1{#1}\fi
\expandafter\ifx\csname bibnamefont\endcsname\relax
  \def\bibnamefont#1{#1}\fi
\expandafter\ifx\csname bibfnamefont\endcsname\relax
  \def\bibfnamefont#1{#1}\fi
\expandafter\ifx\csname citenamefont\endcsname\relax
  \def\citenamefont#1{#1}\fi
\expandafter\ifx\csname url\endcsname\relax
  \def\url#1{\texttt{#1}}\fi
\expandafter\ifx\csname urlprefix\endcsname\relax\def\urlprefix{URL
}\fi \providecommand{\bibinfo}[2]{#2}
\providecommand{\eprint}[2][]{\url{#2}}



\bibitem{Voss} Symbolic Examples: Fourier: R.F. Voss, Phys. Rev. Lett. {\bf 68}, 3805 (1992); 
Walsh-Fourier etc.: D. Stoffer, J. Amer. Statist. Assoc. {\bf 86},
461 (1991); D. Stoffer et al. Biometrika {\bf 180}, 611 (1993);
Mutual information: A. M. Fraser
and H. L. Swinney, Phys. Rev. A {\bf33}, 1134 (1986). 

\bibitem{Billingsley} From three fields: P. Billingsley, Ann. Math. Stat. {\bf 32}, 12
(1961); N. Merhav {\small\emph{et al}} IEEE Trans. Inf. Theory {\bf
35}, 1014 (1989); M.J. van der Heyden {\small\emph{et al}} Physica D
{\bf117}, 299 (1997).




%

\bibitem{Herzel94} H. Herzel {\small\emph{$et\, al$}}, Chaos, Solit. Fract. {\bf 4}, 97 (1994).
%

\bibitem{Berger} T. Cover $\&$ J. Thomas, \emph{Elements of Information Theory} (J. Wiley and Sons, NY, 1991); T. Berger, \emph{Rate Distortion Theory} (Prentice-Hall, Englewood Cliffs, NJ,
1971); A. Gersho and R. Gray, \emph{Vector Quantization and Signal
Compression} (Kluwer Academic, Boston, MA 1992).

\bibitem{Schreiber} T. Schreiber, Phys. Rev. Lett. {\bf 85}, 461 (2000) introduces the `transfer entropy'  to reveal  causal links between two time
series. $\R_{n||m}$ reveals causal connection in the same series,
and can be seen as the \emph{self} transfer entropy.

\bibitem{grassbergertending} P. Grassberger, Int. J. Theor. Phys {\bf
25}, 907 (1986); W. Ebeling and G. Nicolis, Europhys. Lett. {\bf
14}, 191 (1991).

\bibitem{Bandt} Ch. Bandt and B. Pompe, 
J. Stat. Phys. {\bf 70}, 967 (1993).






\bibitem{Dugatkin} This can be connected with multiresolution source coding: eg. D.
Dugatkin (2004) Caltech E. Eng. Thesis.





%
%
%


%
%
%

\bibitem{physionet} Databases: www.physionet.org. Fig. 1 Sudden Cardiac Death Holter Fig. 2 the entire MIT-BIH Normal Sinus Rhythm and BIDMC
Congestive Heart Failure; Fibrillation Data from \cite{Costa}. 



\bibitem{Courtemanche}  M. Courtemanche {\small\emph{$et\, al$}}, Am. J. Physiol. {\bf 257}, 
H693 (1989)  extracts transition matrices from arrhythmias but
discards interbeat intervals.  Many papers have applied symbolic
dynamics to hearts but, to my knowledge, exclude $V$ beats and tend
to record the change in beat intervals symbolically eg. J. Kurths
{\small\emph{$et\, al$}}, Chaos {\bf 5}, 88 (1995).

\bibitem{multiscale} 
P. Bernaola-Galvan
{\small\emph{$et\, al$}}, Phys. Rev. Lett.  {\bf 87}, 168105 (2001).






%
%
%


\bibitem{Schulte} 
V. Schulte-Frohlinde {\small\emph{$et\, al$}}, Phys. Rev. E {\bf 66}, 
031901 (2002).

%
%
\bibitem{costacomm} 
A generalization of the coarsening in Costa {\small\emph{et al}}
\cite{Costa}.


\bibitem{Costa} M. Costa {\small\emph{et al}}, Phys. Rev. Lett. {\bf 89}, 68102 (2002).
%
%
%

\end{thebibliography}
\end{document}